# Acoustic scattering mediated single detector optoacoustic tomography

X. Luís Deán-Ben,[1,2,3,*] Ali Özbek,[1,2,3] Hernán López-Schier[4], and Daniel Razansky[1,2,3,*]

[1]Faculty of Medicine and Institute of Pharmacology and Toxicology, University of Zurich, Switzerland
[2]Institute for Biomedical Engineering and Department of Information Technology and Electrical Engineering, ETH Zurich, Switzerland
[3]Institute of Biological and Medical Imaging (IBMI), Technical University of Munich and Helmholtz Center Munich, Germany
[4]Research Unit Sensory Biology and Organogenesis, Helmholtz Center Munich, Neuherberg, Germany
*Corresponding author: xl.deanben@pharma.uzh.ch

**Abstract**

**Optoacoustic image formation is conventionally based upon ultrasound time-of-flight readings from multiple detection positions. Herein, we exploit acoustic scattering to physically encode the position of optical absorbers in the acquired signals, thus reduce the amount of data required to reconstruct an image from a single waveform. This concept is experimentally tested by including a random distribution of scatterers between the sample and an ultrasound detector array. Ultrasound transmission through a randomized scattering medium was calibrated by raster scanning a light-absorbing microparticle across a Cartesian grid. Image reconstruction from a single time-resolved signal was then enabled with a regularized model-based iterative algorithm relying on the calibration signals. The signal compression efficiency is facilitated by the relatively short acquisition time window needed to capture the entire scattered wavefield. The demonstrated feasibility to form an image using a single recorded optoacoustic waveform paves a way to the development of faster and affordable optoacoustic imaging systems.**

Optoacoustic (OA, photoacoustic) imaging has enabled breaking through the light diffusion barrier to map optical contrast (absorption) with high resolution deep into living organisms [1-3]. This is achieved by capitalizing on the low scattering of ultrasound, as compared to that of light, in soft biological tissues. A myriad of OA systems based on single detector scanning as well as simultaneous acquisition of tomographic information using array detectors have been suggested [4]. In all cases, image formation is based on the assumption that ultrasound waves undergo no distortion and propagate with constant velocity across the sample and coupling medium (typically water). Internal reflections and acoustic scattering may severely deteriorate the quality of the rendered images, thus generally have to be avoided [5,6].

A consequence of acoustic reflections and scattering is the appearance of late responses in the collected time-resolved signals, leading to arc-type artefacts and overall distortion in OA images [7]. Such late responses may nevertheless contain useful information. For example, acoustic reflectors have been included in OA tomographic imaging systems to maximize the effective angular coverage and avoid so-called limited-view effects [8,9]. The late parts of OA signals corresponding to reflected waves can in fact be considered as additional signals collected from mirrored locations, thus effectively doubling the amount of information contained in each signal acquisition. In this way, additional information associated to acoustic reflections can effectively be used to reduce the number of signals required for image formation. Ideally, a single recorded waveform would encode the location of multiple absorbers, provided that a sufficient number of reflected waves is acquired.

In this work, we suggest an alternative approach to encode the location of light-absorbing structures in OA signals based on multiple scattering of ultrasound waves. For this, randomly distributed acoustic scatterers in front of a transducer array result in a unique complex propagation path for the ultrasound waves generated at each source location within the effective field of view. As a result, distinct optoacoustic waveforms are generated by absorbers located at different positions. The number of signals required for reconstructing an OA image can then be significantly reduced without considerably extending the acquisition time window.

A lay-out of the experimental system used to test the suggested concept is depicted in Fig. 1a. A full-ring (360°) array of cylindrically-focused transducers was employed. The OA signals were generated by directly illuminating the region of interest (ROI) by a nanosecond pulsed laser at 720 nm wavelength. The OA signals detected by the array elements were digitized at 40 megasamples per second for a time window of 494 samples delayed by 20 μs with respect to the laser pulse. The collected signals were band-pass filtered between 0.5-8 MHz to remove low-frequency offsets and high-frequency noise. A cluster of acoustic scatterers were randomly distributed along a circular ring coaxially-aligned with the array. Specifically, ~300 borosilicate capillary glass tubings with inside and outside diameters of 0.86 and 1.50 mm respectively (Warner Instruments LLC, Hamden, USA) were distributed along an annulus with 16 mm radius and 20 mm thickness. The custom-made array (Imasonics SaS, Voray, France) has a radius of 40 mm and consists of 512 elements with 5 MHz central frequency and >80% detection bandwidth. The dimensions of the elements are 0.37x15 mm².

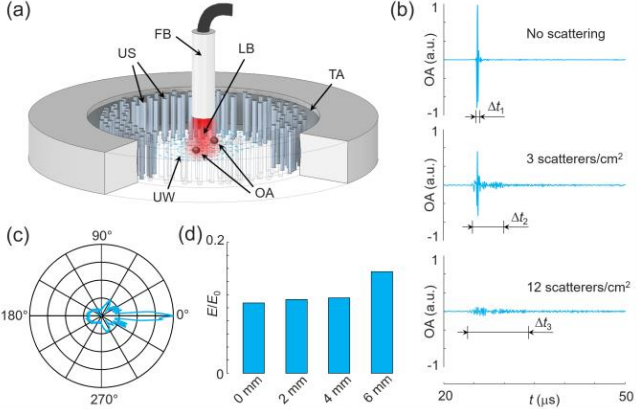

**Figure 1**: Acoustic scattering of optoacoustic waves. (a) Lay-out of the experimental system. TA – transducer array, US – ultrasound scatterers, FB – fiber bundle, LB – laser beam, OA – optical absorbers, UW – ultrasound waves. (b) Collected optoacoustic signals with no scatterers in the propagating path, relatively low and high density of scatterers. (c) Scattered wave directivity for an individual scatterer located at a distance of 16.25 mm from a point absorber. (d) Ratio of the total detected OA signal energy with and without scatterers in the propagating path versus distance of the absorbing microsphere from the center of the transducer array. The signal energy is integrated over all transducer elements and time instants.

The effects of acoustic scattering in the collected OA signals are illustrated in Fig. 1b. For a single 100 μm diameter microsphere absorber (Cospheric LLC, Santa Barbara, USA), the signal detected by one of the array elements with no scatterers in the propagating path is plotted at the top. As expected, the generated signal is confined in time to a short interval corresponding to $\Delta t_1 \sim 1/BW$ centered at $t=d/c$, where BW is the detection bandwidth, $d$ is the distance between the sphere and the sensor and $c$ is the speed of sound. The other two plots show the detected signal when acoustic scatterers are present (Fig. 1a). For the relatively low scattering density of 3 scatterers/cm² the signal extends in time over $\Delta t_2 \sim 5$ μs, yet the part corresponding to direct propagation remains dominant and contains most of the useable information for image reconstruction. Note also some early arriving responses ascribed to a direct acoustic propagation through glass having speed of sound significantly higher than water. The signal detected in the presence of densely distributed 12 scatterers/cm² exhibits a complex pattern spanning $\Delta t_3 \sim 10$ μs and has no dominant peaks. In this case, the location of the point absorber is encoded along the entire recorded interval, thus any given distribution of optical absorbers can in principle be compressed into a single waveform.

We next measured the directivity pattern for an individual scatterer by placing an absorbing microsphere at the center of the transducer array and a glass tubing at a distance of 16.25 mm from it. The relative amplitude of the scattered wave for different angles was estimated by measuring the difference between the OA signals collected by all the array elements with and without the tubing in the propagating path. It is shown that the scattered waves have a dominant forward propagation component. This is expected considering that the effective dimension of each scatterer corresponds to $\sim 5\lambda_a$ ($\lambda_a$ being the acoustic wavelength at the central frequency of the detection array), which falls into the Mie scattering regime. Forward propagation is essential to minimize the loss of energy due to transmission through the scattering medium. Fig. 1d shows the ratio of the total detected OA signal energy for all array elements with (E) and without ($E_0$) scatterers present in the propagating path. For our detection configuration approximately 10% of the OA signal energy is preserved after adding scattering. This value is increased for OA sources located away from the array's center, suggesting that cylindrical focusing of the detection elements contributes to the energy collection efficiency.

Image reconstruction in the presence of acoustic scattering implies establishing a model linking the initial OA pressure (proportional to the optical absorption) to the collected pressure waveforms. Similarly to the time-domain model-based OA reconstruction approaches [10,11], one may assume that the absorbed energy is confined within the region of interest (ROI) and that the acquired OA signals $p(r,t)$ correspond to a linear superposition of the signals $p_i(r,t)$ for each pixel belonging to the image grid covering such ROI, i.e.,

$$p(r,t) = \sum_i h_i p_i(r,t), \quad (1)$$

where $h_i$ is the absorbed optical energy for the *i*-th pixel. For ultrasound waves propagating through a uniform non-scattering medium, $p_i(r,t)$ can be estimated from the OA forward model, e.g. by interpolating light absorption values between the pixel positions [11]. When acoustic scattering takes place, physical modelling of $p_i(r,t)$ implies accurate knowledge of the position, shape and acoustic properties of the scatterers, which may turn too complicated. Instead, $p_i(r,t)$ can be experimentally measured by moving an OA source across a grid of points and collecting the corresponding responses. For this purpose, a 100 μm polyethylene microsphere was scanned with 75 μm steps in the horizontal and vertical directions over a ROI of 4.5x4.5 mm². For the experimentally measured signals, Eq. (1) represents a linear model that can be expressed in a matrix form via

$$\mathbf{p} = \mathbf{M}_p \mathbf{H}, \quad (2)$$

where **p** is a column vector containing the signal(s) for a set of positions and instants and **H** is a vector containing the absorbed energy in each pixel of the scanning grid. The columns of matrix $\mathbf{M}_p$ represent the signal(s) for each position of the scanned particle. More generally, the model in Eq. (2) can be expanded into

$$\mathbf{s} = \mathbf{M}_s \mathbf{H}, \quad (3)$$

where **s** represents any linear combination of signals included in matrix $\mathbf{M}_s$. The performance of the model in Eq. (3) for image reconstruction from a reduced number of signals was first tested by considering a sparse distribution of absorbers. For this, the microsphere was scanned at random grid points within the ROI not included in the calibration grid and the signals were added up. Additionally, the OA signals from all transducer elements were superimposed to form a unique time-resolved signal expressed in a vector form as $\mathbf{s}_m$. Image reconstruction was based upon numerical inversion of Eq. (3), i.e.,

$$\mathbf{H}_{sol} = \mathrm{argmin}_\mathbf{H}\{\|\mathbf{s}_m - \mathbf{M}_s \mathbf{H}\|_2^2 + \lambda^2 \|\mathbf{L}\mathbf{H}\|_l^2\}, \quad (4)$$

where the parameter λ allows weighting the regularization term $\|\mathbf{LH}\|_l^2$, with **L** being an arbitrary matrix. The reconstruction results for an individual microsphere, taking **L** as the identity matrix, indicate that it was possible to accurately reconstruct an image of a point absorber with an OA waveform detected at a single position (Fig. 2a). The full-width at half maximum (FWHM) of the reconstructed sphere when using the L2 norm is ~150 μm, as indicated in the profile in Fig. 2b, which matches the expected in-plane spatial resolution of the array for the scattering-free case [12]. The profile in Fig. 2b for the L1 norm regularization has a significantly lower FWHM. This is attributed to the sparsity condition and must not necessarily be ascribed to the achievable resolution, although L1-based regularization has been shown to enhance the spatial resolution beyond the acoustic diffraction barrier [13]. Accurate reconstructions are similarly rendered when multiple microspheres are present in the ROI (Figs. 2c and 2d), yet resulting in higher noise levels due to reduced sparsity in the images which hampers image reconstruction from a single waveform.

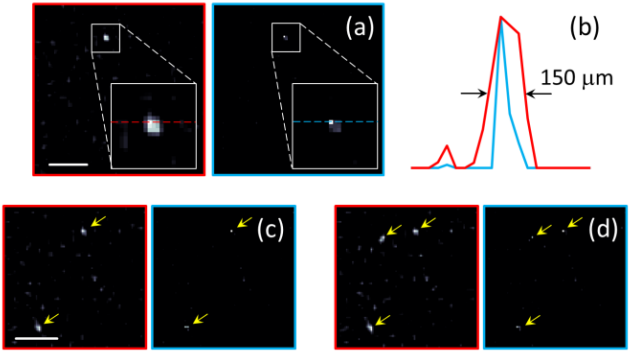

**Figure 2**: Experimental validation of the scattering mediated single detector optoacoustic tomography. (a) Optoacoustic images of a microsphere reconstructed with a single detected waveform. (b) One-dimensional profiles along the reconstructed images shown in (a). Images reconstructed for 2 and 3 microspheres are shown in (c) and (d), respectively. The yellow arrows indicate the position of the spheres. Red and blue squares correspond to inversions done with either L2 or L1 norm (l=2 or l=1 in Eq. (4)). Scalebars – 1mm.

The imaging performance of the suggested approach was further tested by imaging a wild-type zebrafish larva 5 dpf *post mortem*. Image reconstruction was performed considering a total variation regularization term, i.e., Eq. (4) was modified to

$$\mathbf{H}_{sol} = \mathrm{argmin}_{\mathbf{H}}\{\|\mathbf{s}_m - \mathbf{M}_s \mathbf{H}\|_2^2 + \lambda^2 TV(\mathbf{H})\}. \quad (5)$$

Reconstruction made from a signal corresponding to the sum of all 512 signals of the array renders the fish in the correct position, although its shape is distorted (Fig. 3a). Clearly, the image quality improves when reconstructing with 16 signals corresponding to the sum of alternating channels of the array (Fig. 3b). As a reference, Fig. 3c shows the image reconstructed with all the 512 array elements with no scatterers in the ultrasound propagating path and Fig. 3a shows a bright field microscopic image of the larva, where relatively large light absorbing structures are labelled. For the reconstruction of the image in Fig. 3c, a conventional model-matrix $\mathbf{M}_s$ was built as described elsewhere [14]. Even though the reference image exhibits the best reconstruction quality, most structures can be clearly identified in images reconstructed with a significantly lower number of signals.

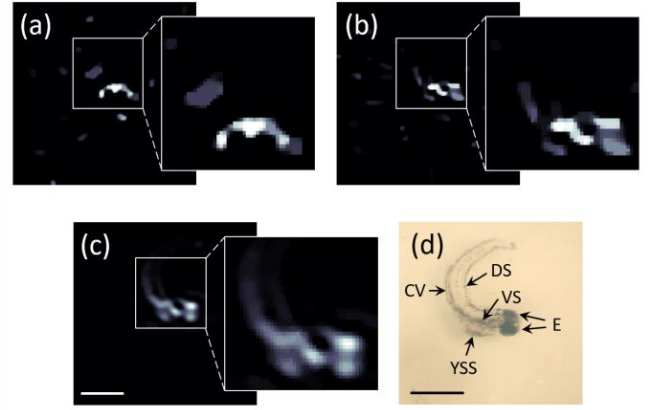

**Figure 3**: Imaging of 5 days-post-fertilization zebrafish larva *post mortem*. (a) Optoacoustic image of the larva obtained with a single integrated waveform. (b) Optoacoustic image of the larva obtained from 16 individual signals. (c) Optoacoustic image of the larva obtained with all 512 signals from the transducer array with no acoustic scattering in the ultrasound propagating path. (d) Bright field microscopy image of the larva. E – eyes, YSS – yolk sac stripe, VS – ventral strip, DS – dorsal stripe, CV – caudal vein. Scalebars – 1mm.

Minimization of the amount of data required for OA image formation may contribute to reducing costs or, alternatively, speeding up acquisitions. Herein, we have demonstrated the basic feasibility to "physically" encode the optical absorption distribution in a defined ROI by capitalizing on the complex propagation of ultrasound waves in a scattering medium. While several time-of-flight readings are normally required to trilaterate position of a given OA source, this information can alternatively be carried via multiple acoustic scattering events. Thus, the suggested approach represents a sort of compressed data acquisition methodology. In a conventional OA imaging scenario, temporal resolution is ultimately limited by the time it takes for all the generated signals to leave the ROI. While this ultimate limit is unattainable with the suggested approach due to the extended acquisition window caused by multiple scattering events, such window is much shorter than that required for alternative approaches using acoustic reflectors [8].

A potential drawback of the suggested methodology is the limited data sparsity, which is essential for high quality OA image reconstruction. We have shown that a single absorber can be very accurately reconstructed with a single waveform, which turns more challenging for multiple sources. Yet, individual flowing absorbers have been previously used in localization OA tomography (LOT) to break through the acoustic diffraction barrier [15,16]. Therefore, compressed acquisition of signals may find applicability for super-resolution imaging of vascular structures. Furthermore, an alternative method based on a plastic acoustic mask placed in front of an ultrasound sensor has been recently suggested for compressed acquisition of signals and reconstruction of sparse images in pulse-echo ultrasound [17]. A distribution of acoustic scatterers similar to the one employed here can thus serve the same purpose while also facilitating hybridization of ultrasonography with OA imaging. Overall, the demonstrated feasibility to form an image with a single OA waveform paves a way to the development of faster and affordable OA imaging systems.

## Acknowledgments

This study was partially supported by the European Research Council Consolidator grant ERC-2015-CoG-682379.